# Analysis on the Study of QoS-Aware Web Services Discovery

T. Rajendran and Dr. P. Balasubramanie

**Abstract**— Web service technology has gained more important role in developing distributed applications and systems on the Internet. Rapid growth of published Web services makes their discovery more and more difficult. There exist many web services which exhibit similar functional characteristics. It is imperative to provide service consumers with facilities for selecting required web services according to their non-functional characteristics or QoS. The QoS-based web service discovery mechanisms will play an essential role in SOA, as e-Business applications want to use services that most accurately meet their requirements. However, representing and storing the values of QoS attributes are problematic, as the current UDDI was not designed to accommodate these emerging requirements. To solve the problems of storing QoS in UDDI and aggregating QoS values using the tModel approach. The aim is to study these approaches and other existing QoS tModel representation for their efficiency and consistency in service discovery. This paper discusses a broad range of research issues such as web service discovery or web service selection based on QoS in the E-Business domain.

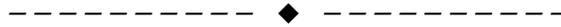

## 1 INTRODUCTION

SERVICE-Oriented Architecture (SOA) supports software to be composed from services dynamically. The SOA is "an architecture that represents software functionality as discoverable services on the network" [4]. SOA provides methods for systems development and integration where systems group functionality around business processes and package these as interoperable services. An SOA infrastructure allows different applications to exchange data with one another as they participate in business processes. Web Services form the core of e-business and hence, have experienced a rapid development in the past few years. Web Services, which are based on XML-based open standards, promise the interoperability of various applications running on heterogeneous platforms. They enable dynamic connections and automation of business processes within and across enterprises for enterprise application integration and business-to-business integration. Building on the ubiquitous and lightweight standards that are supported by major software vendors, Web services enable application integration via the publishing of application's functionality as services, as well as location and invocation of services over the Internet.

Web Services are based on a collection of standards and protocols that allow us to make processing requests to remote systems by speaking a common, non-proprietary language and using common transport protocols such as HTTP and SMTP. The dynamic e-business vision calls for a seamless integration of business processes, applications, and Web services over the Internet. Web service technology enables e-business and e-commerce to become a reality. It has become a competitive tool of companies by reducing cost through fast, effective, and reliable services to customers, suppliers, and partners over the Internet. It enables more efficient business operations via the Web and enhances business opportunities to companies. With the widespread proliferation of Web services, quality of service (QoS) will become a significant factor in distinguishing the success of service providers. Due to the dynamic and unpredictable nature of the Web, providing the acceptable QoS is really a challenging task. Applications with very different characteristics and requirements compete for scarce network resources. Changes in traffic patterns, denial-of-service attacks and the effects of infrastructure failures, low performance of Web protocols, and security issues over the Web create a need for Internet QoS standards. Often, unresolved QoS issues cause critical transactional applications to suffer from unacceptable levels of performance degradation. With the growing popularity and an extensive deployment of Web services, the consumers of Web services will rather want a means to distinguish QoS-aware Web services from QoS-unaware Web services. According to the specification designed by the World Wide Consortium (W3C), QoS requirements for Web services include the following ten major requirements: performance, reliability, scalability, robustness, accuracy, integrity, accessibility, availability, interoperability and security. Given that the QoS-aware Web services can carry a higher business value than other Web services, the Web service providers and requesters would benefit from this kind of services. Web services are new forms of

• T.Rajendran is an Assistant Professor cum Research Scholar, Department of Computer Science & Engineering, SNS College of Technology, Coimbatore, Tamilnadu and India.
• Dr.P.Balasubramanie is a Professor, Department of Computer Science & Engineering, Kongu Engineering College, Perundurai, Tamilnadu and India.

Internet software which can be invoked using standard Internet protocols. Web Services, as it is defined by the World Wide Web Consortium (W3C), is a software system designed to support interoperable machine-to-machine interaction over a network. Web services interact with each other, fulfilling tasks and requests that, in turn, carry out parts of complex transactions or workflows.

UDDI is a platform-independent, XML-based registry for businesses worldwide to list them on the Internet. UDDI is an open industry initiative, enabling businesses to publish service listings and discover each other and define how the services or software applications interact over the Internet. UDDI defines a registry for service providers to publish their services. UDDI is like a registry rather than like a repository. A registry contains only reference information. WSDL is used to describe a web service's capabilities and the interface to invoke it. A WSDL document is self-describing so that a service consumer can examine the functionality of the web service at runtime and generate corresponding code to automatically invoke the service. The interface definition (WSDL) specifies the syntactic signature for a service but does not specify any semantics or non-functional aspects.All these standards are XML-based (eXtensible Markup Language), which allows applications to interact with each other over networks, no matter what languages and platforms they are using. The two features, self-description and language- / platform-independence, distinguish web services from other distributed computing technologies, like CORBA and DCOM.

Fundamentally, standards like WSDL can support service providers describe their services' functionality through standard interfaces and advertise them in some UDDI registries. When a service request is issued, available Web service descriptions and the service request description is matched together in order to find the services that can provide expected functionality (the matching step). However, Web service functional descriptions are not sufficient for service discovery process. There are several reasons for this fact. Firstly, a key advantage of Web service technology is to enable Web services to be dynamically and automatically discovered and selected at runtime. In this case, an automatic mechanism is needed to support a system determine the best services to be chosen. Secondly, as more and more Web services are created by many providers and vendors, there is often the case where a number of Web services can satisfy functional requirements of a service request. Those reasons lead to the issue of ranking and selecting of the best Web services for a request among a list of candidate Web services which can provide similar functionality for the request. Primarily, QoS information is used for computing the quality degree of candidate Web services. Such QoS information can be performance (in terms of response time, latency…), availability, accessibility, security, etc. [14]. These QoS information have substantial impacts on user's expectation and experience of using a Web service. Therefore they can be used as a main factor to distinguish quality of Web services. A Web service with highest ranking value will be selected. This ranking step is often performed after the above matching step. Because different service providers and requesters may use different languages [5], [6], [7], [8] and models [9], [10], [11],[12], [13] for QoS advertisements and requirements, it is necessary to find a way for a system understanding different QoS concepts in QoS descriptions. Besides that, different domains and applications may require different QoS properties; therefore we need a more efficient and flexible method to express QoS information. A study of Web service discovery system is needed to explore existing techniques and to highlight the advantages and disadvantages of each system.

The rest of the article is as follows. The section 2 introduces the background of Web service model, the Web services description languages and the UDDI registry with QoS. Section 3 provides the QoS requirements for web services. Section 4 describes the area of web services discovery. The section 5 presents the issues related to Web services Discovery with QoS, followed by the conclusion in the section 6.

## 2 WEB SERVICE MODEL

In a Web service model, a service provider offers Web services which provide functions or business operations which can be deployed over the Internet, in the hope that they will be invoked by partners or customers; a Web service requester describes requirements in order to locate service providers. Publishing, binding, and discovering Web services are three major tasks in the model. Discovery is the process of finding Web services provider locations which satisfy specific requirements. Web services are useless if they cannot be discovered. So, discovery is the most important task in the Web service model. The Web service model in Fig. 1 shows the interaction between a service requester, service providers, and a service discovery system.

1. The service providers offer Web services which provide functions or business operations. They are created by companies or organizations. In order to be invoked, the Web services must be described. This will facilitate discovery and composition. WSDL or service profile of semantic Web service is used to carry out this function.
2. The Web service requester describes requirements in order to locate service providers. Service requesters usually contain a description of the Web service, though it is not a Web service which can run on the Internet. The requirements are usually described by WSDL, service template or service profile.
3. The Web service discovery or service registry is a broker that provides registry and search functions. The service providers advertise their service information in the discovery system. This information will be stored in the registry and will be searched when there is a request from service requester. UDDI is used as a registry standard for Web service.

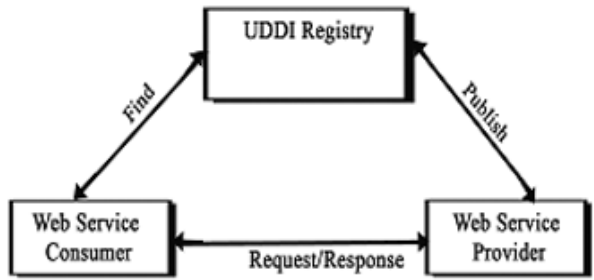

Fig. 1. Web Service Model

The above three components interact with each other via publishing, discovery, and binding operations. These operations are elaborated upon as follows:

1. Publish: the Web service providers publish their service information through the discovery system for requesters to discover. Through the publishing operation, the Web service provider stores the service description in the discovery system.

2. Discovery: the Web service requesters retrieve service providers from the service registry. Based on service descriptions, which describes the requirements of the Web service requesters, the discovery system will output a list of Web service providers which satisfy the requirements.

3. Bind: After discovering, the discovery system provides a number of Web service providers. The Web service requester invokes these Web service providers. The binding occurs at runtime. The Web service requesters and Web service providers will communicate via SOAP protocol which is an XML based protocol for Web service exchange information. Web Service Description Languages

## 2.1 Web Service Description Languages

WSDL [3] is a standard to describe how to access a Web service and what operations (methods) it performs. The WSDL is an XML-based language, which specifies a Web service by defining messages that provide an abstract definition of the data being transmitted and operations that a Web service provides to transmit the messages. WSDL provides a function-centric description of Web services covering inputs, outputs and exception handling. It is also used to locate Web services when a service requester and provider use WSDL to describe the service. A WSDL document defines a Web service using four major elements, namely, port, message, types, and binding. An elaboration of these elements is as follows:

1. WSDL Ports: The <portType> is the most important element in WSDL. It defines a Web service with operations that can be performed and messages that are involved. The <portType> element is similar to a function library (or a module, or a class) in a traditional programming language.

2. WSDL Messages: The <message> element de-

fines an abstract, typed definition of the data being communicated. Each message can consist of one or more logical parts like parameters of a function call in a traditional programming language.

3. WSDL Types: The <types> element defines a container for data type definitions that are relevant for the exchanged messages. For maximum interoperability and platform neutrality, WSDL uses XML Schema syntax to define data types.

4. WSDL Bindings: The <binding> element defines a concrete protocol and data format specification for a particular port type.

## 2.2 The UDDI Registry

UDDI was originally proposed as a core Web service standard. UDDI creates a standard interoperable platform that enables companies and applications to quickly, easily and dynamically find and use Web services over the Internet. It is designed to be interrogated by SOAP messages and to provide access to Web Services Description Language documents describing the protocol bindings and message formats required to interact with the web services listed in its directory. A UDDI registry is a directory for storing information about Web services. A service provider makes its services available to public users by publishing information about the service in a UDDI registry. Individuals and businesses can then locate the services by searching public and private registries. For example, airlines can publish their fare services to a UDDI registry. Travel agencies then use the UDDI registry to locate Web services provided by different airlines, and to communicate with the service that best meets their requirements. The information about Web services in a UDDI registry includes a description of the business and organizations that provide the services, a description of a service's business function, and a description of the technical interfaces to access and manage those services [21].

The UDDI protocol is another XML-based building block of the Web services stack along with SOAP and WSDL. UDDI supplies an infrastructure for systematically addressing needs such as discovery, manageability and security of Web services beyond what is the simple organization of their interactions. By addressing integration, coordination and flexibility issues of service-oriented systems, UDDI plays an important role within the service-oriented approach to enterprise software design. UDDI guarantees flexibility with respect both to the dynamic run-time changes that occur during the life-cycle of web services and to the evolution of web application requirements. UDDI help drive better code reuse and developer productivity. The UDDI specifications define a registry service for Web services and for other electronic and non-electronic services. Service providers can use UDDI to advertise the services they offer. Service consumer can use UDDI to discover services that suites their requirements and to obtain the service metadata needed to consume those services.

## 2.3 Storage of QoS Information in the UDDI Registry

As a current feature in the UDDI registry, the tModel is used to describe the technical information for services. A tModel consists of a key, a name, an optional description and a Uniform Resource Locator (URL) which points to a place where details about the actual concept represented by the tModel can be found [22]. tModels play two roles in the current UDDI registries. The primary role of a tModel is to represent a technical specification that is used to describe the Web services. The other role of a tModel is to register categorizations, which provides an extensible mechanism for adding property information to a UDDI registry. Blum [2, 15] proposes that the categorization tModels in UDDI registries can be used to provide QoS information on bindingTemplates. In the proposal, a tModel for quality of service information for the binding template that represents a Web service deployment is generated to represent quality of service information. Each QoS metric, such as average response time or average throughput is represented by a keyedReference [22], which is a general-purpose structure for a namevalue pair, on the generated tModel. Blum gives an example of the bindingTemplate reference to the tModel with the QoS attribute categories, and an example of the QoS Information tModel, which contains a categoryBag [22], which is a list of name-value pairs specifying QoS metrics. Fig 2 shows the tModel with the QoS Information. This tModel contains a categoryBag that specifies three QoS metrics of Average ResponseTime, Average Throughput and Average Reliability. The tModelKey in each keyedReference is used as a namespace which provides a uniform naming scheme.

```
<tModel tModelKey= "mycompany.com: StockQuoteSer-
vice: PrimaryBinding:QoSInformation" >
<name>QoS Information for Stock Quote Ser-
vice</name>
<overviewDoc>
<overviewURL>
http://<URL describing schema of QoS attributes>
<overviewURL>
<overviewDoc>
<categoryBag>
<keyedReference
tModelKey="uddi:uddi.org:QoS:ResponseTime"
keyName="Average ResponseTime"
keyValue="fast" />
<keyedReference
tModelKey="uddi:uddi.org:QoS:Throughput"
keyName="Average Throughput"
keyValue=">10Mbps" />
<keyedReference
tModelKey="uddi:uddi.org:QoS:Reliability"
keyName="Average Reliability"
keyValue="99.9%" />
</categoryBag>
</tModel>
```

Fig. 2. The tModel with the QoS Information

## 3 QOS REQUIREMENTS FOR WEB SERVICES

The service providers of QoS-aware Web services extend their service descriptions with comprehensible statements pertaining to QoS associated with the entire interfaces or individual components. For a service requestor, these statements are the required QoS from the client's perspective for a service provider, these statements describe what are the QoS levels offered by the server object [32]. Fig. 3 gives the sequence of events and communication between the entities involved for QoS-aware Web services.

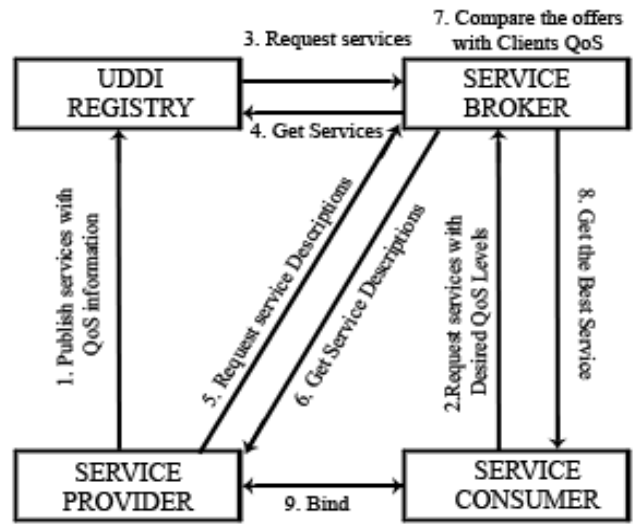

Fig. 3. QoS-aware Web Service operations

The service requestor requests the binding information with the QoS it requires. Depending on the QoS requirements, the broker searches the UDDI for the listed services available and while performing the QoS negotiation by comparing the required and offered QoS, the broker finally determines a QoS that is acceptable for both parties. In this manner, the binding is built and the communication between the service provider and service requestor eventually starts. To support QoS, the developers should be willing to incorporate major design changes to the system, because certain QoS attributes cannot be utilized independently over the existing components. The following sections describe the various QoS requirements for Web services and the techniques to improve those requirements [42].

### 3.1 Performance

The performance of a Web service concerns how fast a Web service request can be processed and serviced. This requirement can be determined by measurements like throughput and latency. Throughput is the measure of the number of requests serviced in specific amount of time. Latency is the amount of delay that is experienced by a client starting from the time when the request is submitted to the time the response information starts to arrive. The request time is the time when the client submits a request to the Web server, and the response time is the time when the Server replies after processing the re-quest. Obviously, the response time and the throughput depend on the workload that the Web server is experiencing at that time. Both latency and throughput can be measured by keeping track of the timestamps at the request time and response times. The latency of any request processing is the difference between the timestamps corresponding to the request time and the response time, and the throughput is given by the total number of requests divided by the elapsed time between the request time and the response time.

### 3.2 Reliability

Reliability requirement measures the quality of a Web service in terms of how it performs, given a specific amount of time and the present network conditions, while maintaining the service quality. It is also determined by the number of failures per day and the medium of delivery. In fact, Web Service Reliability (or WS-Reliability) is a latest specification for open, reliable Web service messaging. WS-Reliability ensures guaranteed delivery of messages, elimination and/or detection of duplicate messages and the right ordering of messages. The WS-Reliability can be embedded into SOAP as an additional extension rather than to a transport level protocol. This specification provides reliability in addition to interoperability, thus allowing communication in a platform and vendor-independent manner. The WS-Reliability specification defines a set of SOAP headers and instructions in SOAP envelopes that manage the message acknowledgements, message ordering etc. WS-Reliability makes asynchronous messaging a healthy choice because of the extra features that it provides like message acknowledgements and message ordering which in turn allow the communicating parties to be more independent of one another [33].

Reliability determines the percentage of the times an event is completed with success. This numeral will provide an opportunity for the service consumers to expect the probability of a failure that can occur during a transaction. The count on failures is based on the number of dropped deliveries, duplicate deliveries, faulty message deliveries, and out-of-order deliveries. An event may either succeed or fail, and there is no middle ground in that issue. Therefore, total number of events will be the number of failures added to the number of successful events.

### 3.3 Scalability

Scalability requirement defines how expandable a Web service can be. Web services are being introduced to new interfaces and techniques everyday and this makes keeping a Web service up-to-date a very important necessity. If the situation demands for more computing capabilities and servicing more requests, the system should be capable of supporting additional systems and newer technologies. The Web service should be able to handle heavy load while making sure that the performance in terms of response time experienced by their clients is not objectionable. The Performance Non-Scalability Likelihood (PNL) metric is a relatively newer technique to predict whether the system is going to be able to withstand the higher loads of traffic without affecting the performance levels. This metric is used to calculate the intensity of the loads at which the system cannot perform without degrading the response time and throughput. The calculation of PNL involves generating potential workloads and studying the behavior of the system which will be similar to how the system would react given such varying workloads. If the system crashes, the engineers will know that it is not scalable enough to accommodate potential future workloads and they could eventually organize an upgrade to the server capability. There can be two states in terms of their behavior. The behavior state can either be acceptable (0) or unacceptable (1). We can also denote the states with relative values in between 0 and 1 (in the interval [0, 1]) instead of constant values to point out the degree at which the system fails to behave acceptably.

### 3.4 Accuracy

Accuracy requirement is a measure of correctness delivered by a Web service. The number of errors generated by a Web service, the number of fatal errors, and the frequency of the situation determine the amount of accuracy for a Web service. Accuracy is defined as the level to which Web services give accurate results for the received requests. Accuracy refers to whether the measurement or metric really measures what we intend to measure. An experiment is conducted to measure the accuracy of the Web service by calculating the standard deviation of the reliability. As the average value of the standard deviation is equal to zero, the measurement is said to be accurate. If the average value of the standard deviation is very high, then the measurement is considered to be not accurate.

### 3.5 Integrity

Integrity requirement assures that any modifications to a Web service are performed in an authorized manner. Data integrity assures that the data is not corrupted during the transfer, and if it corrupted, it assures that there are enough mechanisms in the design that can detect such modifications. Data integrity is an important element to consider, because ignoring it may damage large software modules and create errors that are impossible to trace back. Data integrity is the measure of a Web service's accurate transactional and data delivery abilities. The data messages that are received are verified to see if they have not been modified in transit. This can be done with techniques like checksum calculation or digital signatures. There are a number of tools in the market like SIFT [36] that can collect and monitor the data being sent and received between the communicating parties. These tools can be used to monitor the number of faulty transactions that are unidentified and the data messages that are received but with the checksum or hash that cannot be tallied. Data integrity is a boolean value, meaning that data either has integrity or does not. There is no middle ground or a range that can specify how much integrity the data holds. Data integrity can therefore be calculated as the ratio of successful transactions to the total number of transactions.

$$\text{Integrity} = \frac{\text{Number of successful transactions}}{\text{Total number of transactions}} \quad (1)$$

## 3.6 Availability

Availability requirement is the probability that the Web service is up and in a readily usable state. High availability assures that there is the least amount of system failures or server failures even during the peak times when there is heavy traffic to and from the server and that the given service is available relentlessly at all times. Let us say the "down time" is when a system is not available. As the system is either available or unavailable, the remaining time after subtracting the down time can be termed as the "up time" that means the system is available. As checking upon down time is easier than up time (because down time is smaller than uptime), we can consider to calculate down time to measure the availability. Keeping a tab on the events failed during an operation can possibly give the down time [38].

## 3.7 Accessibility

A Web service's accessibility is the measure of the probability that the client's request to a Web service will be served. Accessibility is typically a measure of the success rate of a service instantiation at a given time. A Web service might not be accessible even though it is still available, because a system may be up and running but might not be able to process a request possibly because of the load it is experiencing. Accessibility in turn depends on how scalable the Web service system design is, because a highly scalable system constantly serves the request irrespective of the volume of the Web service requests. Accessibility is the ratio of the number of successful responses received from the server to the number of requests messages sent by the clients. It can be characterized as the degree of a system at which it is capable of responding to the user invocations of the service. Irrespective of type of acknowledgements received, (either negative/positive or correct/incorrect), accessibility can be calculated as a ratio of number of successful acknowledgements received to the total number of requests sent [38].

$$\text{Accessibility} = \frac{\text{Number of acknowledgements received}}{\text{Total number of requests sent}} \quad (2)$$

## 3.8 Interoperability

Web services are accessed by thousands of clients around the world using different system architectures and different operating systems. Interoperability means that a Web service can be used by any system, irrespective of operating system or system architecture and that accurate and identical result is rendered in any environment. Interoperability is a measure of how flexible the Web service has been created so that the clients do not have to worry about binding to a Web service that cannot be run in their environment. The developmental environment here includes operating system, programming language or hardware type. The inter operability can be calculated as the ratio of the total number of environments the Web service runs to the total number of possible environments that can be used.

$$\text{Interoperability} = \frac{\text{Total number of environments the Web Service runs}}{\text{Total number of possible environments that can be used}} \quad (3)$$

This interoperability value measures the successful execution of the Web service in different environments (such as different operating systems, programming languages, and hardware types). The value closer to 1 indicates higher interoperability which is desirable.

# 4  WEB SERVICES DISCOVERY

Web service discovery is a process of discovering service that most suitable to user's request according to requester's requirement. One of the main challenges in discovering Web services is the fact that service registries do not provide enough query elements for clients to articulate proper service queries that can meet their needs. For example, service registries allow clients to perform simple search queries such as searching by service name, tModel, or business name. However, discovering relevant Web services could not be achieved using simple keyword-based search techniques particularly as Web services proliferate. Furthermore, differentiating Web services from each other using keyword matching techniques is impractical since little textual information is often provided in service discovery interfaces. Discovering web services involves three interrelated phases: 1-matching, 2-assessment, and 3-selection. In phase 1, the service description input by the developer is matched to that of a set of available resources. In phase 2, the result of matching (typically a set of ranked web services) is assessed and filtered by a given set of criteria. In phase 3, services are actually selected for subsequent customizing and combining with others [50]. A discovery service, which could be performed by either a consumer agent or a provider agent, is needed to facilitate the discovery process.

## 4.1 Discovery Mechanisms

As the number of Web services increase, the success of businesses will depend on service discovery. The goal is to find appropriate Web services that match a set of user requirements. A first description of discovery mechanisms for service providers is the *match-making process*. It is the process of finding an appropriate service provider for a service requester through a middle agent [56]. It includes the following general steps: a) Service providers advertise their capabilities to middle agents, b) middle agents store this information, c) a service requester asks a middle agent whether it knows of service providers best matching requested capabilities and d) the middle agent, in order to reply, tries to match the request against the stored advertisements and returns a subset of stored service providers' advertisements. A more up-to-date approach [57] defines the WS Discovery mechanism in a broader sense as "the act of locating a machine-processable description of a WS that may have been previously unknown and that meets certain functional criteria. " It is a service responsible for the process of performing discovery, a logical role, which could be performed by either, the requester agent, the provider agent or some other agent.

WS Discovery mechanisms include a series of registries, indexes, catalogues, agent based and Peer to Peer-P2P solutions. The three leading approaches [25] on how a discovery service should be designed are: a registry, an index, or a peer-to-peer (P2P) system.

**Registry:** A registry is an authoritative, centrally controlled repository of services information. Service providers must publish the information of their services before they are available to consumers. The registry owner decides who has the authority to publish and update services information. A company is not able to publish or update the information of services provided by another company. The registry owner decides what information can be published in the registry. UDDI is an example of this approach. Centralized registries are appropriate in static or a controlled environment where information does not change frequently.

**Index:** An index is a collection of published information by the service providers. It is not authoritative and the information is not centrally controlled. Anyone or any company can create their own index, which collects information of services exposed on the web usually using web spiders. The information in an index could be out of date but can be verified before use. Google is an example of the index approach [25].

**Catalogues:** Web Service catalogues are the dominating technological basis for WS Discovery mechanisms. They are specialized repositories which implement a specification framework as metaschema. In particular, prior to the UDDI standard, organizations lacked a common approach to publish information about their products and web services for their customers and partners. UDDI established the first uniform method that included details for integration of already existing systems and processes between business partners. UDDI allows the enterprises to discover and share information with regard to the web services and other electronic and non-electronic services that are registered in a registry. The requirements include key words, part of the service's name and patience, in order to select the suitable service through the results of the registry. The available search tools are very simple and do not take into consideration any cross-correlations between web services and the qualitative characteristics of each web service, forcing the user to repeat the search from the beginning using new key words. There are three types of information supported by the catalogue. These types included registration of white, yellow and green pages. *White pages* include basic contact information and identifiers such as organization name, address, contact information, and other unique ids. *Yellow pages* describe a web service using different categorizations (taxonomies). This way it is possible to discover a Web Service based upon its category. *Green pages* include the technological information that describes the behaviors and support functions of a Web Service.

**P2P-based Solutions:** P2P can be defined as direct communication or collaboration between computers, where none are simply client or server, but all machines are equals - peers. P2P network is computer network which relies primarily on the computing power and bandwidth of the participants in the network rather than concentrating it in a relatively low number of servers. A pure peer-to-peer network does not have the notion of clients or servers, but only equal peer nodes that simultaneously function as both "clients" and "servers" to the other nodes on the network.

P2P computing provides a de-centralized alternative that allows Web services to discover each other dynamically. Each Web service is a node in a network of peers. At discovery time, a Web service queries its neighbors in search of a suitable Web service. If any one of its neighboring peers matches its requirements, it replies and the query is ended. Otherwise, the query is propagated through the network until a suitable Web service is found or certain termination criteria are reached. P2P architecture is more reliable than registry approach since it does not need a centralized registry, but introduces more performance costs since most of time a node acts as a re-layer of information.

A P2P overlay network provides an infrastructure for routing and data location in a decentralized, self-organized environment in which each peer acts not only as a node providing routing and data location service, but also as a server providing service access. P2P can be considered a complete distributed computing model. Recently proposed P2P systems include CAN [26], Pastry [40] and Chord [41]. The above systems arrange the network of peers to a ring. Nodes are assigned IDs drawn from a global address space. Peers are also assigned a range of keys from the global address space that they are responsible for. Each peer also stores auxiliary information in order to appropriately route key lookups. Usually a key lookup is initiated at a peer. In this case the peer consults its look-up table in order to successfully route the query to the peer that stores the queried key. In the case of Chord routing with the aid of look-up table, simulates binary search on the address space of all peers, thus a request in an $N$ peer network can be routed in $O(\log N)$ time. Chord mainly has been adopted as the overlay P2P network distributed web service architectures. The hosts in the P2P network publish their service descriptions to the overlay, and the users access the up-to-date Web Services.

## 4.2 Manual versus Autonomous Discovery

Depending on who is actually performing the discovery, service discovery could be manual or autonomous. Manual discovery is typically done at design time and involves a human service consumer that uses a discovery service to find services that match its requirements. Autonomous discovery involves a discovery agent to perform this task at design time or run time. One situation in which autonomous discovery is needed is when a service consumer needs to switch to another service because the current service is either no longer available or cannot satisfy its requirements anymore.

## 5 RESEARCH ON QoS-AWARE WEB SERVICES DISCOVERY

Web service and Web service discovery are very active research and development topics. In this section, an overview of major techniques related to our approach is presented. Researchers have proposed various approaches for dynamic web service Discovery. QoS is "a combination of several qualities or properties of a service" [19]. It is a set of non-functional attributes that may influence the quality of the service provided by a Web service [24]. QoS properties describe non-functional aspects of Web services and they are used to evaluate the degree that a Web service meets specified quality requirements in a service request. Basically, they can be categorized into two types: technical quality and managerial quality [43]. The *technical quality* consists of QoS properties related to operational aspects of Web services, such as usability, efficiency, reliability, performance, etc. The *managerial quality* consists of QoS properties used for capturing service management information such as ownership, provider, contract, payment, etc. The QoS requirements for Web services are more important for both service providers and consumers since the number of Web services providing similar functionalities is increasing. Current Web service technologies such as WSDL and UDDI, which are for publishing and discovering Web services, consider only customer functionality requirements and support design time, or static service discovery. Nonfunctional requirements, such as QoS, are not supported by current UDDI registries [18]. Xu et al. [23] shows web service discovery model that contains an extended UDDI to accommodate the QoS information, a reputation management system to build and maintain service reputations and a discovery agent to facilitate service discovery. A service matching, ranking and selection algorithm is also developed, but they did not provide any certification or verification process for QoS in that model.

The web service selection and ranking mechanism uses the QoS broker based architecture [16]. The QoS broker is responsible for selection and ranking of functionally similar web services. The web service selection mechanism [16] ranks the web services based on prospective levels of satisfaction of requester's QoS constraints and preferences. QoS can be used to select and rank the Web services by extending standard SOA [27, 28]. In this architecture, the Web service is selected by matching re-quested QoS property values against the potential Web service QoS property values. In literature, the Web service is selected by taking the requester's average preference for QoS properties [28]. The WS-QoSMan applies an external resource approach in which it uses a tModel called QoSMetrics that contains information to an external reference [20]. This is very similar to the tModel used for pointing to WSDL files. QoSMetrics uses overview URL to point to an XML-based file generated by WS-QoSMan and that contains QoSMetrics for a specific web service.

Hongan Chen et al. [29] presented a description and an implementation of broker-based architecture for controlling QoS of web services. The broker acts as an intermediary third party to make web services selection and QoS negotiation on behalf of the client. Delegation of selection and negotiation raises trustworthiness issues mainly for clients. Performance of the broker is not considered in this approach. Moreover, performance of the broker can be a key to the success of any proposed architecture; if the user does not get a response to his/her request with an acceptable response time, he/she will switch to another provider. Some similar broker based architectures were presented in [30, 31] that focus more on the QoS specification using XML schema, and dynamic QoS mapping between server and network performance. Maximilien and Singh [18] propose an agent framework and ontology for dynamic Web services selection. Service quality can be determined collaboratively by participating service consumers and agents via the agent framework. Service consumers and providers are represented and service-based software applications are dynamically configured by agents. QoS data about different services are collected from agents, aggregated, and then shared by agents. This agent-based framework is implemented in the Web Services Agent Framework (WSAF). A QoS ontology, which captures and defines the most generic quality concepts, is proposed in their paper.

Adam Blum and Fred Carter [1, 2] present four different QoS storing methods in UDDI by utilizing tModels. The first method [1] employs a QoSInformation tModel referring to a QoS file which is not a part of UDDI. The location of the QoS file is stored in the overviewURL of the QoSInformation tModel which is represented in XML, so the detailed QoS information, including QoS values and QoS description, is stored in an XML file. Each bindingTemplate contains one QoSInformation tModel and adds the QoSInformation tModel to the tModelInstanceDetail. The method provides a set of APIs, such as save_business, save_service and save_tModel, to store QoS values. To obtain the desired QoS values requires the built-in APIs (i.e. find_tModel, find_service) to acquire the QoS values in the XML file. The second method [1] creates many different QoS tModels for various QoS information. These categories are added to the bindingTemplates. Each categoryBag has multiple keyedReferences, which represent different types of QoS. Each QoS value is stored in the keyValue of the keyedReference. Apart from the aforementioned API supported by this approach to store the QoS values, two functions are provided (i.e. save_binding and save_tModel) to increase the system flexibility. The third method [1] is similar to the first one by utilizing the QoSInformation tModel, but it contains a bindingTemplate. The categoryBag of QoSInformation tModel has many keyReferences, which represent various types of QoS. The APIs supported by this method is similar to the second approach. The fourth method [2] stores the QoS values in the categoryBag of businessService in UDDI. The method needs save_business, save_tModel and save_service to store QoS values. The method requires find_service to search the Web service according the QoS value. These studies have provided the researchers with better understanding and the usage of UDDI to manage Web Service QoS. However, the QoS information cannot be directly accessed via UDDI APIs, so these approaches either required multiple steps or needed complex queries to locate the appropriate service such that the time required on service discovery would be long. It could lead to the system inefficiency or have scalability problem.

Many researchers work on how to take QoS information for Web services into account in the service discovery process to find services that best meet a customer's requirements. Ran [9] proposes a model for web service discovery with QoS by extending the UDDI model with the QoS information. But service search and selection are still done by human clients. This is not desirable if thousands of services are available for selection. Searching and finding the most suitable service to match the client's functional and QoS needs may be better performed by an automated system module, such as a QoS broker. There are four roles in this proposed model: Web service supplier, Web service consumer, Web service QoS certifier, and the new UDDI registry. The new UDDI registry is a repository of registered Web services with lookup facilities; the new certifier's role is to verify service provider's QoS claims. The proposed new registry differs from the current UDDI model by having information about the functional description of the Web service as well as its associated quality of service registered in the repository. Lookup could be made by functional description of the desired Web service, with the required quality of service attributes as lookup constraints. The Certifier verifies the advertised QoS of a Web service before its registration. The consumer can also verify the advertised QoS with the Certifier before binding to a Web service. This system can prevent service providers from publishing invalid QoS claims during the registration phase, and help consumers to verify the QoS claims to assure satisfactory transactions with the service providers. Although this model incorporates QoS into the UDDI, it does not provide a matching and ranking algorithm, nor does it integrate consumer feedback into service discovery process.

Gouscos et al. [17] proposed a simple approach to dynamic Web services discovery that models Web service management attributes such as QoS and price. They discuss how this simple model can be accommodated and exploited within basic specification standards such as WSDL. The key Web service quality and price attributes are identified and categorized into two groups, static and dynamic. The Price, Promised Service Response Time (SRT) and Promised Probability of Failure (PoF) are considered as static in nature and could be accommodated in the UDDI registry. The actual QoS values that are the actual SRT and PoF are subject to dynamic updates and could be stored either in the UDDI registry or in the WSDL document, or could be inferred at run time through a proposed information broker. The advantage of this model is its low complexity and potential for straightforward implementation over existing standards such as WSLA and WS-Policy specifications.

Fatih Emekci et al. [44] proposed a Web service discovery method that considers both the functionality and the behavior of the Web services, while providing a scalable reputation model for ranking the discovered services. The method operates over a peer-to-peer system, thus avoiding the inherent problems of centralized systems such as scalability, single point of failure and high maintenance cost. Peer-to-peer systems provide a scalable alternative to centralized systems by distributing the data and load among all peers. These systems are decentralized, scalable and self-organizing. Current peer-to-peer systems, such as Napster, Gnutella and KaZaA, are mainly being used for sharing data over the Internet. Locating data and routing messages within these systems are accomplished through a centralized index or flooding. These systems are referred to as unstructured peer-to-peer systems, because the overlay network is constructed in a random manner and there is no restriction on where the data can be stored. Thus search in these systems tends to be inefficient because there is no way to determine which peers in the system are more likely to have certain data. The proposed reputation model allows Web services to rate each other's trust and service quality like online auction sites. Our reputation model is scalable in terms of storage, computation and maintenance cost as it is based on approximate counting. It enables the ranking of the search results according to the trust and service quality ratings of the Web services.

Hunaity and Rashid [45] refines the web service discovery process through designing a new framework that enhances retrieval algorithms by combining syntactic and semantic matching of services. It proposes a new framework for smarter WS discovery that provides clients with QoS information which will enhance the selection process and reduce the failure chances by getting endorsements or recommendations from other services or special agents about each service. The proposed model consists of the basic web service model components (Service Provider, Service Consumer, and UDDI Registry) with one addition, which is the capability to store QoS information using tModel data structure. The model is enhanced with a three agents (Discovery Agent, Service Mediator and Reputation Manager). The service provider will describe the entire functional and non functional attribute in the UDDI directly, or through the service mediator agent. The service mediator agent will handle all communication with registries, bindings, negotiations, voting, requests, and responses for that service. The service consumer can search for a specific service directly in the UDDI or it can interact with its specific discovery agent. This framework does not provide any verification or certification process.

Weifeng Lv and Jianjun Yu [46] uses WSDL-S as Web services semantic description language, and constructs a P2P overlay network for Web services distribution, discovery and invocation supporting locality-preserving, range query, tree lookups and semantic service matching. It also promotes a Web Services discovery system named *pService* and achieve considerable performance comparable to centralized service discovery systems. This new system provides a novel architecture to aggregate and classify massive loose-coupled and heterogeneous services on Web. It enables easier service publishing for service provider; and more quickly, succinctly service discovery, integration and invocation for service requester.

Maximilien and Singh [47] proposed a multi-agent based architecture to select the best service according to the consumers' preferences. Maximilien and Singh describe a system in which proxy agents gather information on services, and also interact with other proxy agents to maximize their information and the conceptual model they use to interact with the services is detailed elsewhere [48]. The proxy agents lie between the service consumer and the service providers. The agents contact a service broker, which contains information about all known services, as well as ratings about its observed QoS. From there, the information is combined with its own historical usage, and the combined knowledge is used to select a service, though the authors do not detail how. The agencies contain data about the interactions between the clients and the services which is used during the Web Services selection process. In his work, trust and reputation are taken into account during the decision process. Their approach divides the QoS attributes into objective and subjective. The former include QoS features such as availability, reliability, and response time. Liu, Ngu, and Zeng [49] consider these features in their proposed approach as well but their major selection criteria is based on the QoS based service selection. They have considered three quality criteria namely execution time, execution duration and reputation for the selection. In addition, execution price, duration, transactions support, compensation and penalty rate are the other criteria. Liu, Ngu, and Zeng [49] suggests an open, fair, and dynamic framework that evaluates the QoS of the available Web Services by using clients' feedback and monitoring. The reasoning mechanism is responsible for the selection of a Web Service at a particular moment of time. In order to distinguish one service from another using the specified criteria, this unit requires a set of instructions that help evaluate each component and choose the most appropriate one respectively. A set of instructions can be seen as a selection technique. The major components of a reasoning mechanism are criteria, model, and selection technique. The model collects information about the participants of the client-server interaction as well as represents it as aggregated measures. Different selection techniques can implement various business logics in order to make a decision. The reasoning mechanism in the approach proposed by Liu, Ngu, and Zeng [49] computes the QoS of the Web Services, ranks them, and selects the most appropriate one.

To perform the selection, the QoS registry in their system takes in data collected from the clients, stores it in a matrix of web service data in which each row represents a web service and each column a QoS parameter, and then performs a number of computations on the data, such as normalization. Clients can then access the registry, and are given a service based on the parameters that the client prefers. The bottleneck of the approach is the dependency on the consumers to give regular feedback about their past experience with the Web Services. The success of this model is based on the clients or the end users and their will to provide the necessary feedback on QoS.

Diego and Maria [51] proposed an extended Web service architecture to support QoS management. In the proposed approach, QoS information derived from policies based on WS-Policy is encapsulated inside QoS Policy (*qosPolicy*) structures stored in UDDI registries. WS-Policy comprises a model and corresponding syntax to specify Web service policies. References to policies can be attached to XML documents, and policies can be applied at different levels, including the process, service and operation levels. UDDI is extended to include QoS policies and the enhanced service description may be used in service discovery. These extensions to UDDI registry and query mechanisms enhance search flexibility. Each element of a *qosPolicy* structure can be associated with a Technical Model (*tModel*) structure, which allows specification, standardization and reuse of QoS-related concepts. Furthermore, the extension allows the use of brokers to facilitate service selection according to functional and non-functional requirements, and monitors to verify QoS attributes. The proposed architecture is currently being integrated with Business Process Management (BPM) technology.

Farnoush et al [52] introduces WSPDS (Web Services Peer-to-peer Discovery Service), a fully decentralized and interoperable discovery service with semantic-level matching capability. The peer-to- peer architecture of the semantic-enabled WSPDS not only satisfies the design requirements for efficient and accurate discovery in distributed environments, but also is compatible with the nature of the Web Services environment as a self-organized federation of peer service-providers without any particular sponsor. With peer-to-peer services, the role of distinct service providers is eliminated. System entities all cooperate to provide a service as a result of group collaboration in a distributed fashion. Entities are peers in functionality and each entity is potentially both a server and a client of the peer-to-peer service; hence, sometimes entities are referred to as servents (i.e., server and client). WSPDS is a distributed discovery service implemented as a cooperative service. A network of WSPDS servents collaborate to resolve discovery queries raised by their peers. To discover a service requested by user, a servent originates a query (enveloped in a SOAP message) in the network of servents. The servents collaborate to propagate the query based on the probabilistic flooding dissemination mechanism. Dissemination of a query is restricted by its TTL. Each servent is composed of two engines, communication engine and local query engine, standing for the two roles that a servent plays. The communication engine provides the interface to user and also represents the servent in the peer-to-peer network of servents. The local query engine receives the queries from the communication engine, queries the local site (where the servent is running) for matching services, and sends responses to the communication engine.

Wang et al. [53] proposed a QoS Ontology language designed for the needs of Web services is defined in the context of WSMO (Web Services Modeling Ontology). It also presents a QoS selection model, as well as associated selection mechanism using an optimum normalization algorithm. However, the selection model is general for individual service selection. Web service composition is not discussed in that work. Zeng et al. [54] discuss a global planning approach for selecting composite Web services. They propose a simple QoS model containing five QoS criteria, namely price, duration, availability, reliability, and reputation. They apply linear programming for solving the selection optimization problem in service composition. However, this approach is often too complex for run-time decisions. Reference [55] proposes software architecture to provide QoS-enabled web services by adding a QoS broker between clients and service providers to discover the QoS aware services in UDDI. However, no detailed information about QoS broker, such as how it is designed and the functionality of it is presented.

Demian et al. [34], explores different types of requester's QoS requirements and a tree model for requester's QoS requirements. It also proposed a QoS broker based web service architecture which facilitates the requester to select a suitable web service based on QoS requirements and preferences. Serhani et al. [39] shows web service architecture which employs an extended UDDI registry to support service selection based on QoS, but only the certification approach is used to verify QoS and no information is provided about the QoS specification. Many of these approaches do not provide guarantees as to the accuracy of the QoS values over time or having up-to-date QoS information.

## 6 CONCLUSION

Making the provision of services QoS-aware is to the advantage of both clients and providers in the e-business domain. This paper presented an analysis and study of Web services discovery with QoS Management systems. The purpose of web services discovery is to select optimal web service for a particular task. QoS plays an important role in Web service discovery in order to evaluate and rank candidate Web services that are able to provide expected functionality. A number of QoS-broker based web service discovery framework have been developed recently. This paper reviews those approaches, analyze their advantages and disadvantages as well as specify open issues that need further research in ongoing works aiming at developing architectures and related algorithms for matching and ranking Web services based on QoS.

## REFERENCES

[1] Adam Blum and Fred Carter, "Representing Web Services Management Information", Available:http://www.oasis-open.org/committees/ download.php/5144/

[2] Adam Blum, "UDDI as an Extended Web Services Registry ", SOA WORLD MAGAZINE, Available: http://webservices.sys-con.com/read/45102.htm

[3] E.A.Walsh, "UDDI, SOAP, and WSDL: The Web Services Specification", Reference Book (1st edition ed.) 0130857262: Pearson Education, 2002.

[4] W.Han, "Integrating Peer-to-Peer into Web Services". Master thesis, University of Saskatchewan, 2006.

[5] S.Frolund and J.Koisten, "QML: A Language for Quality of Service Specification", Hewlett-Packard, http://www.hpl.hp.com/techreports/98/HPL-98-10.htm,1998.

[6] Heiko Ludwig, Alexander Keller, Asit Dan, Richard P. King, Richard Franck, "Web Service Level Agreement (WSLA) Specification", http://www.research.ibm.com/wsla/, 2002.

[7] A.Sahai, A.Durante, and V.Machiraju, "Towards Automated SLA Management for Web Services", HP, http://www.hpl.hp.com/techreports/2001/HPL-2001-310R1.pdf, 2002.

[8] V. Tosic, B. Pagurek, and K. Patel. "WSOL: Web Service Offerings Language". In the book of Web Services, E-Business, and the Semantic Web, LNCS, Springer- Verlag, pp. 57-67, 2003.

[9] S. P. Ran. "A Model for Web Service Discovery with QoS". ACM SIGecom Exchanges, Vol. 4, No. 1, ACM Press, pp. 1–10, 2003.

[10] M. Matinlassi and E. Niemela. "The Impact of Maintainability on Component-based Software Systems". In Proc. of the 29th Euromicro Conference (EUROMICRO.03), Turkey, 2003.

[11] B. Sabata, S. Chatterjee, M. Davis, J. J. Sydir, and T. F. Lawrence. "Taxonomy for QoS Specifications". In Proc. of the 3rd International Workshop on Object-Oriented Real-Time Dependable Systems, USA, 1997.

[12] L. Chung, B. A. Nixon, E. Yu, and J. Mylopoulos. "Non-Functional Requirements in Software Engineering". In Kluwer Academic Publishing, 2000.

[13] S. W. Choi, J. S. Her and S. D. Kim. QoS Metrics for Evaluating Services from the Perspective of Service Providers. In Proc. of the IEEE International Conference on e-Business Engineering, ACM, pp. 622–625, 2007.

[14] J. C. Laprie, B. Randell, and C. Landwehr. "Basic Concepts and Taxonomy of Dependable and Secure Computing". In IEEE Transactions on Dependable & Secure Computing. Vol. 1, No. 1, pp. 11–33, 2004.

[15] A.Blum,"Extending UDDI with Robust Web Services Information". Retrieved from http://searchsoa.techtarget.com/ news/article/0,289142,sid26_gci952129,00.html, 2004.

[16] Demian.A.D'Mello and V.S.Ananthanarayana, "A QoS Model and Selection Mechanism for QoS-Aware Web Services", Proceedings of the International Conference on Data Management (ICDM2008), 2008.

[17] D.Gouscos, M.Kalikakis, and P.Georgiadis, "An Approach to Modeling Web Service QoS and Provision Price". Proceedings. of the 1st International Web Services Quality Workshop - WQW 2003 at WISE 2003, Rome, Italy, pp.1-10, 2003

[18] E.M.Maximilien and M.P.Singh, "A Framework and Ontology for Dynamic Web Services Selection". IEEE Internet Computing, 8(5):84-93, 2004.

[19] D.A.Menascè, "QoS Issues in Web Services", IEEE Internet Computing,6(6):72-75, 2002

[20] Eyhab Al-Masri and Qusay H. Mahmoud, "Discovering the Best Web Service", WWW 2007, Canada, 2007.

[21] UDDI.org, "UDDI Technical White Paper", Retrieved from http://uddi.org/pubs/uddi-tech-wp.pdf

[22] UDDI.org, "UDDI Version 2.03 Data Structure Reference", Retrieved from http://uddi.org/pubs/DataStructure-V2.03-Published-20020719.htm, 2002.

[23] Ziqiang Xu, Patrick Martin, Wendy Powley and Farhana Zulkernine, 2007, "Reputation Enhanced QoS-based Web services Discovery", IEEE International Conference on Web Services (ICWS 2007), IEEE 2007.

[24] W3C Working Group, "QoS for Web Services: Requirements and Possible Approaches". Retrieved from http://www.w3c.or.kr, 2003.

[25] W3C Working Group, "Web Services Architecture, W3C Working Group Note 11 February 2004". Retrieved from http://www.w3.org/TR/wsarch, 2004.

[26] S.Ratnasamy, P.Francis, M.Handley, R.Karp and S.Shenker, "A scalable content-addressable network". Proceedings of ACM SIGCOMM¨01, San Diego, September 2001.

[27] M. A. Serhani, R. Dssouli, A. Hafid, and H. Sahraoui, "A QoS broker based architecture for efficient web service selection", Proceedings of the IEEE International Conference on Web Services (ICWS'05), IEEE 2005.

[28] J. Hu, C. Guo, H. Wang, and P. Zou, "Quality Driven Web Services Selection", Proceedings of the IEEE International Conference on e-Business Engineering (ICEBE'05), IEEE 2005.

[29] Hongan Chen, Tao Yu, Kwei-Jay Lin, "QCWS: an implementation of QoS-capable multimedia web services", IEEE Fifth International Symposium on Multimedia Software Engineering, IEEE 2003.

[30] M. Tian, A. Gramm, T. Naumowicz, H. Ritter, J. Schiller, "A Concept for QoS Integration in Web Services", 4th International Conference on Web Information Systems Engineering, Rome, Italy, 2003.

[31] M. Tian, A. Gramm, H. Ritter, J. Schiller, "Efficient selection and monitoring of QoS-aware web services with the WS-QoS framework", IEEE/WIC/ACM international Conference on Web Intelligence, Beijing, China, 2004.

[32] A.Nagarajan, and A. Mani, "Understanding quality of service for Web services". IBM Developer Works, January 2002.

[33] Sun, "Web Services Reliability (WS-Reliability) Version 1.0: frequently asked questions (FAQ)". Sun Developer Network (SDN), January 2003.

[34] Demian Antony D´ Mello, V.S.Ananthanarayana and T.Santhi, "A QoS Broker Based Architecture for Web Service Selection", Proceedings of IEEE International Conference, 2008.

[35] R.L.Costello, "Web Services Best practice – Summary 3". XML-Dev Technical Articles, January 2002. .

[36] B.Barbash, "Sift 1.5 by service integrity". SOA Web Services Journal, September 2003.

[37] D.Talbot, "Data integrity in Web services". Win_Dev.com, C# help. March, Technical Article, 2004.

[38] OASIS. "Summary of quality model forWeb services (OASIS Web Services Quality Model TC)". OASIS, June 2002.

[39] M.A.Serhani, R.Dssouli, A.Hafid and H.Sahraoui, "A QoS broker based architecture for efficient Web services selection". In Proc. of the IEEE Int'l Conf. on Web Services, IEEE CS, pages 113–120, 2005.

[40] A.Rowstron, P.Druschel, "Pastry: Scalable, decentralized object location and routing for largescale peer-to-peer systems". Proc. of the 18th IFIP/ACM Middleware, Germany. 2001.

[41] I. Stoica, R. Morris, D. Karger, F. Kaashoek, and H. Balakrishnan, "Chord: A Scalable Peer-to-peer Lookup Service for Internet Applications". In Proceedings of ACM SIGCOMM, 2001.

[42] D.U.Weider, B.R.Rachana, Sumana Pingali and Vijaya Kolluri,"Modeling the Measurements of QoS Requirements in Web Service Systems", Simulation,83(1), pp.75-91, January 2007.

[43] J. C. Laprie, B. Randell, and C. Landwehr. "Basic Concepts and Taxonomy of Dependable and Secure Computing". In IEEE Transactions on Dependable & Secure Computing. Vol. 1(1), pp. 11–33, 2004.

[44] Fatih Emekci, Ozgur D. Sahin, Divyakant Agrawal, Amr El Abbadi. "A Peer-to-Peer Framework for Web Service Discovery with Ranking". Proceedings of the IEEE International Conference on Web Services (ICWS'04), 2004.

[45] Mossab Ahmmad Rashid, Hunaity, "Towards an Efficient Quality Based Web Service Discovery Framework". IEEE Congress on Services, 2008.

[46] Weifeng Lv, Jianjun Yu. "pService: Peer-to-Peer based Web Services Discovery and Matching". Second International Conference on Systems and Networks Communications (ICSNC 2007), 2007.

[47] E.M. Maximilien and M.P. Singh. "Towards Autonomic Web Services, Trust and Selection". ICSOC'04 pages 212–221, November 2004.

[48] Maximilien, E.M. and Singh, M.P. "Reputation and endorsement for web services". SIGecom Exch. Volume 3.1, pages 24-31. 2001.

[49] Y. Liu, S. Ngu, and L. Zeng. "QoS Computation and Policing in Dynamic Web Service Selection" WWW2004, May 2004.

[50] N. Kokash, W. Van Den Heuvel, and V. D'Andrea, "Leveraging Web Services Discovery with Customizable Hybrid Matching," University of Trento, July 2006.

[51] Diego Zuquim Guimarães Garcia, Maria Beatriz Felgar de Toledo. "A Web Service Architecture Providing QoS Management". Institute of Computing University of Campinas, São Paulo, Brazil, 2006.

[52] Farnoush Banaei-Kashani, Ching-Chien Chen, and Cyrus Shahabi. "WSPDS: Web Services Peer-to-peer Discovery Service". University of Southern California, Los Angeles, California, 2004.

[53] X. Wang, T. Vitvar, M. Kerrigan, and I. Toma, "A QoS-Aware Selection Model for Semantic Web Services," in Proceedings of ICSOC 2006, pp. 390-401. June 2006.

[54] Zeng L. Z., Benatallah B., A. Ngu H. H., et al. "QoS-aware middleware for Web services composition," IEEE Transaction on Software Engineering, vol. 30, no. 5, pp. 311-327, 2004.

[55] P. Farkas and H. Charaf, "Web Services Planning Concepts", 1st International Workshop on C# and .NET Technologies on Algorithms, Computer Graphics, Visualization, Distributed and WEB Computing, Feb. 2003.

[56] K.Decker, K.Sycara and M.Williamson, "Middle-Agents for the Internet". Proc. 15th IJCAI. Nagoya, Japan. pp.578-583, 1997.

[57] D.Booth, H.Haas, F.McCabe, E.Newcomer, M.Champion, C.Ferris and D.Orchard, "Web Services Architecture". W3C WG Note. http://www.w3.org/TR/ws-arch/, 2004.

**T.Rajendran** is a PhD Scholar in the Department of Computer Science and Engineering at Kongu Engineering College, affiliated to Anna University, Coimbatore, Tamilnadu, and India. His research interest includes Web Service discovery with QoS and Web Technology. He has obtained ME degree in Computer Science and Engineering. He is a life member of ISTE. He has published more than 18 articles in International/ National Journals/Conferences.

**Dr.P.Balasubramanie** has been awarded Junior Research Fellowship (JRF) by CSIR in the year 1990. He has completed his PhD degree in theoretical computer science in the year 1996 in Anna University Chennai. Currently he is a professor in the department of Computer science and Engineering in Kongu Engineering College, Perundurai, Tamilnadu, and India. He has guided 5 PhD scholars and guiding 20 scholars Under Anna University. He has published more than 57 articles in International/ National Journals. He has authored 6 books with the reputed publishers.